
\magnification=1200

\headline={\ifnum\pageno>1{\ifodd\pageno\rightheadline 
\else\leftheadline\fi}\fi}
\def\leftheadline{\headrm\hfil K.C.\ HANNABUSS\hfil}
\def\rightheadline{\headrm\hfil QUANTUM HALL FLUIDS, $\ldots$\hfil}

\font\sym=lasy10
\font\bigbf=cmbx10 scaled\magstep1
\font\smallrm=cmr7 scaled\magstep1
\font\headrm=cmr7 
\font\smallbf=cmbx7 scaled\magstep1

\def\frac#1#2{{#1\over #2}}

\def\aut{{\rm Aut}}
\def\real{{\bf R}}
\def\complex{{\bf C}}
\def\alg{{\cal A}}
\def\ideal{{\cal I}}
\def\hilb{{\cal H}}
\def\tr{{\rm tr}}
\def\rk{{\rm rk}}
\def\bra#1{{\langle #1|}}
\def\ket#1{{|#1\rangle}}
\def\ip#1#2{{\langle #1|#2\rangle}}
\def\conj#1{\overline{#1}}
\def\thm#1{{\bigskip\noindent{\bf #1}.}}
\def\pf{\medskip\noindent{\it Proof. }}
\def\qed{{\sym\char50}}
\def\endpf{\quad\qed}
\def\ref#1{\advance\count30 by 1
\medskip\item{\the\count30.} #1}
\def\sect#1{\advance\count32 by 1
\bigskip{\bf \item{\the\count32.} #1} 
\medskip\noindent}

\count32=0
\count30=0

\centerline{\bigbf QUANTUM HALL FLUIDS, LAUGHLIN WAVE FUNCTIONS,}

\centerline{\bigbf AND IDEALS IN THE WEYL ALGEBRA}

\bigskip
\centerline{K.C.\ HANNABUSS}

\medskip
\centerline{\it Balliol College, Oxford, OX1 3BJ, England.} 

\bigskip
\noindent{{\bf Mathematics Subject Classifications (2000):}
58B34, 35Q58, 53D55, 22E46,  81SO5}

\noindent{{\bf Keywords:} noncommutative geometry, quantum Hall effect, 
Laughlin wave function, Calogero--Moser, Weyl algebra.}

\bigskip\noindent
{\smallbf Abstract.}
{\smallrm
It is known that noncommutative fluids used to model the Fractional Quantum 
Hall effect give Calogero--Moser systems. The group-theoretic description 
of these as reductions of free motion on type A Lie algebras leads directly 
to Laughlin wave functions. The Calogero--Moser models also parametrise the 
right ideals of the Weyl algebra, which can be regarded as labelling sources 
in the fluid.}

\sect{Introduction}
In this letter we draw attention to some connections between recent 
work on noncommutative fluids and the fractional quantum Hall effect, 
the Calogero--Sutherland and Calogero--Moser integrable models,  and 
mathematical investigations of ideals in the Weyl algebra.
The main result is a group theoretical derivation of Laughlin wave functions 
by considering the Calogero--Moser model as the reduction of free motion on 
the Lie algebra $su(N)$.

The transverse magnetic field $B$ which generates the Hall effect forces 
the replacement of the translation group $E =\real^2$ of the plane sample by 
the magnetic translations, which form a projective representation of $E$ 
with a multiplier determined by the field.
Since translations in orthogonal directions no longer commute this suggests 
the use of the methods of noncommutative geometry, and Bellissard and Connes
were able to interpret the conductance as a Chern character determined by 
pairing a cyclic 2-cocycle for the twisted group algebra of $E$ determined 
by the multiplier with the projection onto the Fermi level and thence to 
obtain a very clear mathematical understanding of the integer Hall effect, 
[3,8].

To some extent this succeeded because the integer quantum Hall effect can 
be explained within one electron theory, whilst the fractional quantum Hall 
effect seems to be an essentially many body problem.
Laughlin suggested early on that one should model the many 
electrons as an incompressible two-dimensional fluid, and there has been much 
work exploiting that idea [2,26].
Susskind suggested combining this approach with noncommutative 
geometry and looking at incompressible fluid flow in the noncommutative plane,
[29], a suggestion soon taken up by others [27,17].
Douglas and Nekrasov have provided a useful review of noncommutative field 
theories in general [12].

The main result on Laughlin wave functions is derived in Section 4. It is 
preceded by brief accounts of quantum fluids and the Calogero type models in 
Sections 2 and 3.
The final section describes the link between sources in the fluid and ideals 
in the Weyl algebra.

\vfill\eject
\sect{Quantum Hall fluids}
Two dimensional incompressible fluids and the noncommutative plane share
essentially the same symmetry group (at least as a discrete group, though not 
as an algebraic group, see [4]).
To be more precise the noncommutative plane has a coordinate algebra generated
by elements $y^1$ and $y^2$ which satisfy a commutation relation of the form 
$$[y^j,y^k] = \theta^{jk} 1,$$
where $\theta^{jk}$ are the components of the symplectic form $\theta$ on $E$ 
whose exponential defines a chosen multiplier on $E$.
(We shall use natural identifications to identify $\theta$ as a non-degenerate
antisymmetric bilinear form on the vector space $E$ with a symplectic form 
on $E$ considered as a differentiable manifold.
To preserve flexibility we do not insist that this be the same as 
the multiplier for magnetic translations introduced above, though, since 
$\wedge^2E$ is only one-dimensional, it will necessarily be a power of the 
magnetic multiplier.)
We can then form the Weyl algebra of Schwartz functions on $E$ with Moyal 
multiplication determined by $\theta$.
The automorphism group of the Weyl algebra consists of bijective maps from 
$E$ to $E$ which preserve the Schwartz functions, so that they should be 
differentiable (with some asymptotic conditions), and also preserve 
the Moyal product forcing them to preserve the symplectic form $\theta$.
The uniqueness of $\theta$ up to multiples, means that these diffeomorphisms 
are the same as those preserving the area two-form.
On the other hand the flow of an incompressible fluid is described precisely 
by the volume-preserving (or in two dimensions, area preserving) 
diffeomorphisms of the region in which it is contained.
(This description of fluid flow is clearly laid out in [1].)

One clear advantage of the noncommutative plane over the commutative one is 
that every automorphism (respectively, every derivation) is generalised inner, 
that is, can be implemented as conjugation by (respectively, commutator with) 
an element of the multiplier algebra. 
(For the algebra of Schwartz functions this just consists of tempered 
distributions whose products with any Schwartz functions are again Schwartz 
functions [Cor].)
Some obvious derivations are actually inner.
For example, (as noted already by Born, Jordan, and Dirac [5,11]) the 
map $f \mapsto [y^j,f]$ is easily checked to give 
$\theta^{jk}\partial f/\partial y^k$.
So introducing the dual form $\theta_{jk}$, we have
$$\partial f/\partial y^k = \theta_{kj}[y^j,f].$$
Any fluid motion adds to $y^j$ a displacement term, and the commutator with 
the displaced fluid coordinate can be interpreted as a connection $\nabla$ 
obtained by adding a function of the noncommutative coordinates to the 
partial derivative $\partial_k = \partial/\partial y^k$.
More precisely, we write $\nabla = d+a$ for a connection for a finite rank 
projective module of the Weyl algebra (the analogue of a vector bundle over 
the noncommutative space), and this gives
$$[\nabla_k,f]  = [\partial_k,f] + [a_k,f]
= \theta_{kj}[y^j,f] + [a_k,f]
= \theta_{kj}[y^j + \theta^{jl}a_l,f],$$
so that the connection replaces $y^j$ by $z^j = y^j + \theta^{jl}a_l$.
It is worth noting that the curvature of this connection is 
$$[\nabla_j,\nabla_k] 
= \theta_{jr}\theta_{ks}[y^r + \theta^{rl}a_l,y^s + \theta^{sm}a_m]
= \partial_ja_k - \partial_ka_j + [a_j,a_k] + \theta_{jk}.$$
Due to the Moyal product, $[a_j,a_k]$ can contribute even for abelian gauge 
groups, and the last term arises from $[y^r,y^s]$.

The resulting reinterpretation of the fluid flow as a theory of connections 
has been used in the physics literature to recast the theory as a 
noncommutative Chern--Simons theory [29,27].
The gauge group is often said to be the whole group of unitary inner 
automorphisms, but this ignores the important asymptotic conditions.
These have recently been discussed by Harvey [15], who argues that, rather 
than the unitary group of the Hilbert space, as is often assumed, the 
gauge group of the noncommutative theory should only consist of unitary 
functions with asymptotic values in the group of unitaries $U$ such that 
$U-1$ is asymptotically a compact operator.
In any case the gauge group action allows us to gauge away any incompressible 
fluid motion (or at least any with the same asymptotic conditions).
It cannot, however, remove sources or vortices which exist within the fluid.
One is left with a physical picture in which the filled or partially filled 
Landau levels of the system are modelled as a fluid, but there are in addition
some additional particles, quasi-particles, or sources.
The gauge theory enables us to ignore the fluid and concentrate on these.

\sect{The Calogero--Moser model}
Susskind started investigation of the resulting noncommutative Chern--Simons 
theory by looking at matrix versions.
Although there are no finite-dimensional solutions to the commutation relation 
$[y^j,y^k] = \theta^{jk}1$, an heuristic argument suggests that if the fluid 
contains a source of strength $q$ in a state $\psi$ then the commutation 
relation is amended to $[y^j,y^k] = \theta^{jk}(1-q\ket{\psi}\bra{\psi})$,
which (taking traces) has $N\times N$ matrix solutions if and only if 
$q\|\psi\|^2 =N$.
Polychronakos was able to reach a similar conclusion in a more direct way by 
consideration of a theory containing additional source or droplet fields 
transforming with the fundamental representation of $U(N)$.

It is known that there are solutions of this modified commutation relation   
and that they describe solutions of the Calogero--Moser model [25,19].
We briefly outline the latter approach.
This starts by considering the cotangent bundle 
$T^*u(N)\sim u(N)\oplus u(N)^*$ 
of the Lie algebra $u(N)$ of the unitary group $U(N)$.
We identify $u(N)$ and $u(N)^*$ with the self-adjoint $N\times N$ matrices, 
and write $(X,P)$ for a typical element of $T^*u(N)\sim u(N)\oplus u(N)^*$.
The dual $u(N)^*$ can also be identified with $u(N)$ using the trace 
(Hilbert--Schmidt) inner product.
The unitary group acts by conjugation.
This is a symplectic map (with respect to the standard symplectic structure on
the cotangent bundle $T^*u(N)$, and we readily calculate that the moment map 
$\mu: T^*u(N) \to u(N)^* \cong u(N)$ is given by $\mu(X,P) = [P,X]$.
Identifying $P$ and $X$ with $y^1$ and $y^2$, the unitary equivalence class of 
matrix solutions of the modified commutation relation can therefore be found 
by symplectic reduction to 
$\mu^{-1}(\theta^{12}(1-q\ket{\psi}\bra{\psi})/U_\psi(N)$,
where $U_\psi(N)$ is the stabiliser (little group) of the projection onto 
$\psi$.

On the other hand each unitary orbit can be parametrised as follows.
The eigenvalues $\{x_1,x_2,\ldots,x_N\}$ of $X$ are unitary invariants.
Let $\{e_1,e_2,\ldots,e_N\}$ be the corresponding eigenvectors, and 
$p_j = \ip{e_j}{Pe_j}$ the diagonal entries of $P$.
The commutation relation tells us that if $\psi = q^{-\frac12}\sum e_j$
(which is consistent with the normalisation condition) then the $x_j$ must be 
distinct and that, the off-diagonal entries ($j\neq k$) of $P$ are given by 
$\ip{e_j}{Pe_k} = \theta^{jk}(x_j-x_k)^{-1}$.
This enables us to parametrise the orbit by 
$(x_1,\ldots, x_N, p_1,\ldots,p_N)$.
These coordinates are unique up to permutations of the $N$ indices, and this 
is the only vestige of the original unitary symmetry which survives the 
reduction.
The Hamiltonian $\frac12\tr(P^2)$ in the unreduced space reduces to the 
Calogero--Moser Hamiltonian
$$\frac12\sum_{j=1}^N p_j^2 - \frac12\sum_{j<k} \theta_{jk}^2(x_j-x_k)^{-2}.$$
(A similar analysis on $T^*U(N)$ gives the Calogero--Sutherland model for 
particles on the circle instead of the real line.
This case is particularly interesting because a rigorous second quantisation 
has recently been described by Carey and Langmann [7].)

It is a standard mathematical trick that such a reduction can also be 
achieved by reduction of the larger space $T^*u(N)\times\complex^N$. 
The second term is equipped with the imaginary part of the inner product as 
symplectic form.
The moment map for the conjugation action on $T^*u(N)$ combined with the 
natural action of $U(N)$ on $V=\complex^N$ is 
$\mu(X,P,\psi) = [X,P] - \ket{\psi}\bra{\psi}$,
and reduction at $\theta^{12} 1$, gives us the same reduced manifold as before.
(The stabiliser is now $U(N)$ whose action is enough to remove the extra 
degrees of freedom which we had introduced.) 
This construction actually recovers the extra fields which Polychronakos 
introduced for physical reasons [26,27]. 
(Although introduced as fields, the constraint equations force determine them 
up to a vector in $V$.) 

An even more subtle variant appears in [30], where everything is complexified,
and one considers the $GL(N)$ action on $T^*(gl(N)\oplus V)$.
The moment map on an element 
$(X,P,v,w)\in T^*(gl(N)\oplus V) \cong gl(N)\oplus gl(N)^* \oplus
V\oplus V^*$ is 
$$\mu(X,P,v,w) = [X,P]-\ket{v}\bra{w},$$
and reduction gives the complex Calogero--Moser equation.
(As Wilson notes the complex version allows collisions between particles, but 
the unreduced space desingularises the effect of these collisions [30].) 

\sect{The quantised Calogero--Sutherland and Calogero--Moser models}
The quantised version of reduction is to exponentiate the point of the dual 
Lie algebra at which the system is reduced, to a (linear) character of the 
stabiliser, and then to restrict attention to the Hilbert subspace on which 
it acts by this character.
We work with the second version of the construction on 
$T^*u(N)\times V$.
In fact, since the adjoint action on $u(N)$ is trivial on the orthogonal 
complement of $su(N)$ we restrict to that
It is thus convenient to use the Schr\"odinger Hilbert space $L^2(su(N))$ for 
the first factor and the Fock space 
$\bigoplus_{r=1}^\infty\otimes_S^rV$ for the second factor.
We first remove the factor in front of the vector part (to agree with the unit
factor in the moment map) by rescaling to $X=q^{-\frac12}y^1$, 
$P=q^{-\frac12}y^2$ and then reduce at $\theta^{12}/q$.
Analysis of this reduction goes back to [27,17], but we shall analyse it 
within the standard setting of compact group representation theory, leaving 
physical considerations till the end.

\thm{PROPOSITION}
The exponential of $q^{-1}\theta^{12}$ defines a character of $U(N)$ 
only when $\theta^{12} = kq$ for integral $k$.
Thus the filling factor $\nu = q/\theta^{12} = 1/k$ is the reciprocal of 
an integer.
The corresponding reduced space describes fermionic or bosonic particles 
for odd or even $k$ respectively.
For a given value of $k$ the reduced space involves only the component 
$\oplus_S^{kN}V$ in the Fock space of $V$.

\pf
Any character of $U(N)$ must be lifted from its quotient by the commutator 
subgroup $SU(N)$, and it is easy to check that they are just (integral) powers 
of the determinant: $U\mapsto \det(U)^k$.
Exponentiating $\theta^{12}/q$ tells us that the relevant character of $U(N)$
is given by $k=\theta^{12}/q$, and shows, in particular, that 
$q^{-1}\theta^{12}=k$ is an integer.
As we noted earlier the permutation group is the vestige of the unitary gauge 
group which survives reduction.
A permutation matrix has determinant $1$ or $-1$ as the permutation is odd or 
even, so the quantised particles are bosons or fermions according to whether 
the integer is even or odd.  
Physically $\nu=q/\theta^{12}$ is interpreted as the filling factor, so       
we have shown that it is the reciprocal of an integer, and for fermionic 
behaviour it must be an odd integer.

To find the isotypic component transforming with this character, we first 
consider just the scalars $\lambda 1\in U(N)$.
These have a trivial conjugation action on $su(N)$, and so the representation 
on $L^2(su(N))$ is trivial.
On the other hand the action on $\otimes_S^rV$ is by $\lambda^r$.
Noting that $\det(\lambda 1)^k = \lambda^{Nk}$, we see that 
$r= Nk = N/\nu$ is permitted.
\endpf

\bigskip
This result holds equally well for the Calogero--Sutherland model where 
the space $L^2(su(N))$ is replaced by $L^2(SU(N))$, whilst the mathematics 
becomes particularly transparent in that case so we shall start with that. 
We shall show, by considering the action of $SU(N)$, that for integer 
values of $k$ there is indeed a non-trivial reduced space, which can be 
described very precisely in terms of Laughlin type wave functions. 
The principle of the argument is already apparent for $k=1$ so we first deal 
with that.

\thm{THEOREM}
The part of $L^2(SU(N)) \otimes \otimes_S^{N}V$  which transforms with 
$\det(U)$ under the action of $U\in U(N)$ has a distinguished cyclic vector 
$\Delta_0$ (considered as a $\otimes_S^{N}V$-valued function of
$Z \in SU(N)$) defined by 
$$\ip{v^{(N)}}{\Delta_0(Z)} 
= \ip{{Z^*}^{N-1}v\wedge\ldots \wedge{Z^*}^2v\wedge Z^*v\wedge v}
{\epsilon},$$
where $v^{(N)}= v\otimes v\otimes\ldots\otimes v\in \otimes_S^NV$, 
and $\epsilon$ is a fixed unit vector in $\bigwedge^NV$, and it is 
spanned by functions of the form 
$\Delta_{\bf k}(Z) = \chi_{\bf k}(Z)\Delta_0(Z)$, where $\chi_{\bf k}$ is the 
character of the irreducible representation of  $SU(N)$ with highest weight 
${\bf k}$.

\pf
We have already dealt with the action of multiples of the identity.
(Moreover, every unitary matrix can be written as $\lambda U$ with $U\in SU(N)$ 
and $\lambda\in \complex$, unique up to multiplying $\lambda$ by an $N$-th 
root of unity, and in $\otimes_S^NV$  this ambiguity has no effect.)

By the Peter--Weyl theorem $L^2(SU(N))$ decomposes under the left and right 
actions of $SU(N)\times SU(N)$ as the direct sum of terms $D\otimes D^*$ 
where $D$ ranges over the irreducible representations of $SU(N)$.
We are interested only in the adjoint action of the diagonal subgroup 
$SU(N) \subset SU(N)\times SU(N)$ and seek those $D$ for which 
$D\otimes D^*\otimes\otimes_S^NV$ carries a trivial $SU(N)$ representation.
This means that we need a non-trivial intertwining operator $T$ from $D$ to 
$D\otimes\otimes_S^NV$.
The Littlewood--Richardson rule [22,23] tells us that this is possible if and 
only if $D$ has highest weight 
${\bf k} = (k_1,k_2,\ldots,k_{N-1})$ with 
$k_1 > k_2 > \ldots >k_{N-1} >0$.
Moreover, when this constraint is 
satisfied there is, up to multiples, a unique intertwining operator.
The minimum highest weight permitted by this constraint has $k_j = N-j$, so 
that ${\bf k} = (N-1,N-2,\ldots,1) = \rho$ is precisely half the sum of the 
positive roots.

The corresponding representation $D_\rho$ on a space $\hilb_\rho$ can also be 
realised as the highest weight irreducible summand in the tensor product 
representation $D_\Lambda$ on 
$\Lambda = V\otimes\wedge^2V\otimes\ldots\otimes\wedge^{N-1}V$ given 
by the natural action on $V$.
It is easy to write down an intertwining operator from 
$\Lambda$ to $\Lambda \otimes \otimes_S^NV$.
For $x = x_1\otimes x_2\otimes\ldots\otimes x_{N-1}$ and 
$y = y_1\otimes y_2\otimes\ldots\otimes y_{N-1}$, with $x_j,y_j\in \wedge^jV$,
$\epsilon\in \wedge^NV$, and $v\in V$ we set
$$\ip{y\otimes v^{(N)}}{T(x)} = \ip{v}{x_1}\ip{v\wedge y_1}{x_2}\ldots
\ip{v\wedge y_{N-2})}{x_{N-1}}\ip{v\wedge y_{N-1}}{\epsilon}.$$ 

The Peter--Weyl Theorem associates to the operator $T_v$ defined by
$$\ip{y}{T_v(x)} =  \ip{y\otimes v^{(N)}}{T(x)},$$
the function  on $G$ given by $Z \mapsto \tr(T_vD_\rho(Z)^*)$.
Another application of the Littlewood--Richardson rule shows that the only 
contribution to the trace comes from the highest weight components, so we may 
as well replace $D_\rho$ by $D_\Lambda$, which is much easier to compute as
the sum of terms $\ip{x}{T_v(x)}$ with $x$ in an orthonormal basis, that is 
each $x_j$ in an orthonormal basis of $\wedge^jV$.
The definitions give  
$$\eqalign{\ip{x}{T_vD_\Lambda(Z)^*x} 
&=  \ip{x\otimes v^{(N)}}{TD_\Lambda(Z)^*x}\cr
&= \ip{v\wedge x_{N-1}}{\epsilon}\ip{v}{Z^*x_1}\ip{v\wedge x_1}{\wedge^2Z^*x_2}
\ldots \ip{v\wedge x_{N-2}}{\wedge^{N-1}Z^*x_{N-1}}.\cr}$$
Summing over an orthonormal basis of $x_1$ we find that
$$\sum_{x_1}\ip{v}{Z^*x_1}\ip{v\wedge x_1}{\wedge^2Z^*x_2} 
= \sum_{x_1}\ip{Zv}{x_1}\ip{Zv\wedge Zx_1}{x_2} 
= \ip{Zv\wedge Z^2v}{x_2}.$$
Multiplying this by $\ip{Zv\wedge \wedge^2Zx_2}{x_3}$ 
and summing over orthonormal $x_2$ gives 
$$ \sum_{x_2}\ip{Zv\wedge Z^2v}{x_2}\ip{Zv\wedge \wedge^2Zx_2}{x_3}
= \ip{Zv\wedge Z^2v\wedge Z^3v}{x_3},$$ 
and inductively the sequence collapses down to 
$\ip{Zv\wedge Z^2v\wedge\ldots\wedge Z^Nv}{\epsilon}$.
Since $\epsilon$ is invariant under $Z\in SU(N)$ this can also be written as
$\ip{v\wedge Zv\wedge\ldots\wedge Z^{N-1}v}{\epsilon}$, or as
$\ip{Z^{-(N-1)}v\wedge \ldots\wedge Z^{-1}v\wedge v}{\epsilon}$.
This last version extends from $SU(N)$ to $U(N)$,  since the action of 
$U\in U(N)$ gives
$$\eqalign{\ip{v^{(N)}}{U^{(N)}\Delta_0(U^{-1}ZU)}
&=\ip{(U^{-1}v)^{(N)}}{\Delta_0(U^{-1}ZU)}\cr
&= \ip{(U^{-1}ZU)^{-(N-1)}U^{-1}v\wedge \ldots \wedge U^{-1}v}{\epsilon}\cr
&= \ip{U^{-1}Z^{-(N-1)}v\wedge\ldots \wedge U^{-1}v}{\epsilon}\cr
&= \ip{Z^{-(N-1)}v\wedge\ldots \wedge v}{\wedge^NU\epsilon}\cr
&= \det(U)\ip{(v^{(N)}}{\Delta_0(Z)},\cr}$$
showing that $\Delta_0$ is simply multiplied by $\det(U)$ and so in the 
required subspace.
Replacing $\Delta_0(Z)$ by its $k$-th symmetric tensor power gives similarly a
$\otimes_S^{Nk}V$ valued function which transforms with $\det(U)^k$.

Almost all unitary matrices $Z$ have eigenvectors $e_1,\ldots,e_N$ with 
distinct eigenvalues of modulus 1
$z_1,\ldots,z_N$ we may take $\epsilon = e_1\wedge\ldots\wedge e_N$, and then 
$$\eqalign{
\ip{v^{(N)}}{\Delta_0(Z)} 
&= \ip{Z^{-(N-1)}v\wedge \ldots \wedge v}{\epsilon}\cr
&= \det(\ip{v}{Z^{N-s}e_r})\cr
&=\det(z_r^{N-s}\ip{v}{e_r})\cr 
&= \Delta({\bf z})\prod_{r=1}^N\ip{v}{e_r},\cr}$$
where $\Delta({\bf z}) = \prod_{r<s}(z_r-z_s)$.
Thus $\Delta_0(A) = \Delta({\bf z})\widetilde{e}$, where we have written 
$\widetilde{e}$ for the symmetric tensor product of the eigenvectors of $Z$.
Although $Z$ has distinct eigenvalues on all but a closed submanifold of lower 
dimension, there remains some subtlety when $A$ passes through such a 
submanifold, since the eigenvectors can then be permuted.
However, conjugation invariance means that this problem can be avoided by 
selecting a Cartan subgroup and arranging the ordering in advance.
This shows that, unlike the scalar function $\Delta$, the vector-valued 
function $\Delta_0$ can be extended from a Cartan subgroup to the whole group.

Now the general function in $L^2(SU(N))\otimes\otimes_S^NV$ can be written in 
the form $\chi(Z)\Delta_0(Z)$, where $\chi(U^{-1}ZU) = \chi(Z)$ in order that 
it transform with $\det(U)$, that is $\chi$ is a central function.
The central functions are spanned by the irreducible characters 
$\chi_{\bf k}$,  so that the functions $\Delta_{\bf k} = \chi_{\bf k}\Delta_0$ 
span the space transforming with $\det(U)$.
Using Weyl's character formula and earlier notation, we could write 
$\chi_{\bf k}(Z) = \det(z_r^{k_s+N-s})/\Delta({\bf z})$, so that
$$\ip{v^{(N)}}{\Delta_{\bf k}(Z) }
= \ip{Z^{-k_1-(N-1)}v\wedge Z^{-k_2-(N-2)}v\wedge\ldots\wedge v}{\epsilon}
\widetilde{e}.$$
The inner product on the reduced space involves only integration over a 
cross-section of adjoint orbits. It is therefore sufficient to integrate 
over a Cartan subgroup $H$ of diagonal matrices.
Thus
$$\ip{\Delta_{\bf k}}{\Delta_{\bf l}} 
= \int_H \ip{\Delta_{\bf k}(h)}{\Delta_{\bf l}(h)}\,dh,$$
where the second inner product is that in $\otimes_S^{N}V$.
Substituting  the formula for $\Delta_{\bf k}$ gives
$$\ip{\Delta_{\bf k}}{\Delta_{\bf l}} 
= \int_H \conj{\chi}_{\bf k}(h)\chi_{\bf l}(h)|\Delta(h)|^2\,dh
= \int_{SU(N)} \conj{\chi}_{\bf k}(Z)\chi_{\bf l}(Z)\,dZ,$$
the normal inner product on $L^2(SU(N))$.
\endpf

\bigskip
This result has an easy extension to the Calogero--Moser model.

\thm{COROLLARY}
The part of $L^2(su(N)) \otimes \otimes_S^{N}V$  which transforms with 
$\det(U)$ under the action of $U\in U(N)$ has a distinguished cyclic vector 
$\delta_0$ (considered as a $\otimes_S^{N}V$-valued function of $Z$ in 
$su(N)$) (identified with self-adjoint matrices) defined by 
$$\ip{v^{(N)}}{\delta_0(Z)} 
= \ip{{Z}^{N-1}v\wedge\ldots \wedge{Z}^2v\wedge Zv\wedge v}
{\epsilon},$$
where $v^{(N)}= v\otimes v\otimes\ldots\otimes v\in \otimes_S^NV$, 
and $\epsilon$ is a fixed unit vector in $\bigwedge^NV$, and it is 
spanned by functions of the form 
$\delta_{\bf k}(Z) = \chi_{\bf k}(Z)\delta_0(Z)$, where $\chi_{\bf k}$ is 
the extension to general matrices of the polynomial giving the
character of the irreducible representation of  $SU(N)$ with highest weight 
${\bf k}$.

\pf
There are two ways to derive this result.
The direct method is to note that $T^*u(N)$ can be identified with the 
complexified Lie algebra $u(N)_\complex$, on which there are mutually 
commuting actions of $U(N)$ by left and right multiplication.
Howe duality [16] on the quantised space $L^2(u(N))$ shows that it decomposes 
exactly as in the Peter--Weyl theorem.
Alternatively, one may use the Cayley transform 
$$Z \mapsto U_Z=(1-iZ)(1+iZ)^{-1}[\det(1+iZ)/\det(1-iZ)]^{-1/N}\in SU(N),$$ 
valid on all but a null set, to identify $L^2(su(N))$ with $L^2(SU(N))$.
(As noted before the ambiguity in the root of the determinant has no effect in 
this case, and the map commutes with the adjoint actions of $SU(N)$ on itself 
and its Lie algebra.)
Then we simply apply the previous arguments to obtain a distinguished wave 
function $\delta_0^\prime\in L^2(su(N))\otimes \otimes_S^NV$ 
such that, for $Z\in su(N)$ 
$$\ip{v^{(N)}}{\delta_0^\prime(Z)} 
= \ip{U_Z^{-(N-1)}v\wedge\ldots \wedge U_Z^{-2}v\wedge U_Z^{-1}v\wedge v}
{\epsilon},$$
Replacing $v$ by $(1+iZ)v$ and recalling that $\wedge^N(1+iZ) = \det(1+iZ)$
we obtain
$$\eqalign{\det(1-iZ)^{(N-1)}&\ip{v^{(N)}}{\delta_0^\prime(Z)}\cr
& = \ip{(1-iZ)^{-(N-1)}v\wedge(1-iZ)^{-(N-2)}(1+iZ)v\wedge
\ldots \wedge (1+iZ)^{N-1}v}{\epsilon},\cr}$$
which reduces to a numerical multiple of 
$\ip{Z^{N-1}v\wedge Z^{N-2}v\wedge\ldots \wedge v}{\epsilon}$,
essentially the same function as for the Calogero--Sutherland model.
The rest of the result follows as before, and the isotypic subspace 
is spanned by vectors 
$$\chi_{{\bf k}}({\bf z})\delta^\prime(Z)
= \det(z_r^{k_s+N-s})\widetilde{e}.$$

The Cayley transform has Radon--Nikodym derivative $|\det(1+iZ)|^{-2}$, so 
giving a Cauchy measure on $su(N)$.
However, because we used only the adjoint action of $SU(N)$ on $L^2(SU(N)$ 
rather than the action of $SU(N)\times SU(N)$ one may change the measure on the 
Cartan subalgebra $h$ which provides a cross-section of the orbits.
\endpf

\bigskip
In the physics literature it is customary to choose a Gaussian measure
$\exp[-\frac12\omega\tr(Z^2)]$ appropriate to changing the free motion on 
$su(N)$ to an oscillator motion with Hamiltonian $\frac12\tr(P^2 + 
\omega^2X^2)$.
(Polychronakos motivates this normalisation factor by adding an appropriate 
harmonic oscillator potential to the action.)
Since $\tr(Z^2)$ is invariant under the adjoint action the transformation 
properties are not compromised.
In fact for such oscillatory motion it is more natural to do a Bargmann 
transform  from $L^2(SU(N))$ to the space of holomorphic functions of 
$Z \in su(N)_\complex$ which are square-integrable with respect to the measure 
$\exp[-\frac12\omega\tr(Z^*Z)]$.
Alternatively one multiplies the wave functions by 
$\exp[-\frac14\omega\tr(Z^*Z)]$ rather than by $\det(1-iZ)^{-(N-1)}$, 
giving precisely the Laughlin wave function (of non-self-adjoint $Z$)
$$\ip{v^{(N)}}{\delta_0(Z)} =  
\ip{{Z^*}^{N-1}v\wedge {Z^*}^{N-2}v\wedge\ldots \wedge v}{\epsilon}
\exp[-\frac14\omega\tr(Z^*Z)].$$

Our functions $\Delta_0$ and $\delta_0$, derived by these group-theoretic 
arguments, are essentially the same as the ground state which appears in [17], 
but our factorisation makes clearer the analogy with the Laughlin ground state.
They also appear in correlation functions for the second quantised 
Calogero--Sutherland model [7].
It should be noted however, that the second quantised Calogero--Sutherland 
model is not simply a quantisation of this model, since it combines all 
particle numbers and so all values of $N$ together.

\sect{Sources and ideals in the Weyl algebra}
A very different approach to the theory can be made by returning to the idea 
of sources in the fluid.
The interpretation of diffeomorphisms as a gauge theory meant that we could 
effectively remove regular incompressible fluid motions by a gauge 
transformation.
The connection can, however, still encode for sources.
In the commutative theory of incompressible flow these $N$ sources would have 
positions, which could be defined by an ideal of index $N$ in the coordinate 
ring $\complex[x,y]$ which vanishes at precisely those $N$ points.
(In the spirit of algebraic geometry we work now in the complex plane.)

The obvious generalisation to noncommutative fluids would be to study ideals 
in the noncommutative ring $\alg_1 = \complex[y^1,y^2]$, where 
$[y^1,y^2] = \theta 1$.
We need now to be more precise and work with right ideals, $\ideal$, but clearly 
$\ideal$ cannot have finite index as the quotient $\alg_1/\ideal$ would then provide a 
finite dimensional representation of the commutation relations.
However, Cannings and Holland have classified the right ideals in $\alg_1$, 
by constructing an isomorphism to a Grassmannian of subspaces of 
the rational functions $\complex(z)$, [6].
Berest and Wilson, [4], note that the same Grassmannian arises in Wilson's 
work on the link between the Calogero--Moser model and the KP hierarchy [30], 
where it is shown to be isomorphic to the disjoint union of spaces 
$${\cal C}_N = \{(X,Y)\in gl(N,\complex): \rk([X,Y]-1) \leq 1\},$$ 
(where $\rk$ denotes the rank of a matrix).
(Berest and Wilson note their later discovery that the unpublished thesis of 
Kouakou [20] had anticipated parametrisation by an integer $N$.)
The unitary equivalence classes of the pairs $(X,Y)$ give precisely the 
reduction studied earlier and, as already noted, can be identified with the 
(complex) $N$-particle Calogero--Moser system.
Each right ideal $\ideal$ is associated under the isomorphisms with a point in one 
of the ${\cal C}_N$, so that although $\ideal$ has infinite index, it still 
defines a system of $N$ points, and, moreover, these have a natural 
Calogero--Moser dynamics.

One of the major theorems of Berest and Wilson, says that the natural 
action of the automorphism group $\aut(\alg_1)$ on the right ideals pulls back 
to a transitive action on each individual ${\cal C}_N$.
As they observe, this is surprising because it is much stronger than the 
corresponding result in the commutative case, where the action cannot 
be transitive since it cannot move between configurations where the points are 
distinct and those in which some points are coincident.
As often happens quantisation desingularises.

Finally picking up on another observation of Berest and Wilson, we note that 
the space which reduced to the Calogero--Moser model has an obvious 
generalisation to 
$T^*(gl(N)\oplus M_{N,r})$, where $M_{N,r}$ denotes the $N\times r$ matrices 
(over $\complex$).
Retaining the earlier notation but with $v\in M_{N,r}$ and $w\in M_{r,n}$,
the corresponding moment map is given by $\mu(x,p,v,w) = [X,Y]-vw^*$.
However, the map $j:(X,Y,v,w)\mapsto (Y^*,-X^*,w^*, -v^*)$
which anticommutes with $i$ and has square $-1$, provides  
a hyperk\"ahler structure from which one can obtain an additional  
(real) moment map, 
$$\mu_\real(X,Y,v,w) = \frac12\left([X,X^*]+[Y,Y^*] + vv^*-w^*w\right).$$
The hyperk\"ahler reduction agrees with the previous reduction and gives the same space ${\cal C}_N(r)$:
$$\mu^{-1}(1)\cap\mu_\real^{-1}(0)/U(N) = \mu^{-1}(1)/GL(N,\complex).$$
Multiplication by $(i+j)/\sqrt{2}$ gives an isomorphism with another reduction
$${\cal D}_N(r)= \mu^{-1}(0)\cap\mu_\real^{-1}(1)/U(N),$$
which describes the noncommutative ADHM data for the construction of an 
$r$-instanton solutions of noncommutative self-dual Yang--Mills theory, [24].
This links the subject with other recent work on noncommutative instantons,
[9,28,18,12,14], and shows that the Calogero--Moser model can 
be regarded as a one instanton solution.
This also opens the way to the description of more general filling fractions 
of the form $r/k$.

There remains the question of whether the noncommutative incompressible fluid
model can be derived directly from quantum field theory, and here we note that
Fr\"ohlich and Studer found a limiting commutative Chern--Simons action
for non-relativistic charged spinning electrons [13], and recently Lieb, 
Seiringer and Yngvason have derived the Gross--Pitaevskii energy functional 
directly from a bosonic field theory [21], providing hope that this can be 
done.
(It is interesting that in the Lieb--Seiringer--Yngvason limit the part of 
the two body interaction potential which survives is precisely the inverse 
square part as appears in the Calogero--Moser model.)

\bigskip\noindent
{\bf Acknowledgements}

\medskip\noindent
The first draft of this paper was written during a visit to Adelaide at the 
invitation of V. Mathai, to whom the author is grateful for many 
useful conversations on various aspects of noncommutative geometry and the 
quantum Hall effect. He would also like to acknowledge helpful discussions 
on various occasions with Alan Carey and with Edwin Langmann.

\vfill\eject   
{\bf References}

\ref
Arnold, V.: {Sur la g\'eometrie diff\'erentielle des groupes de Lie 
de dimension infinie et ses applications \`a 
l'hydrodynamique des fluides parfaits}, {\it Ann. Inst. Fourier} {\bf 16} 
(1966), 320-361.
\ref
Bahcall, S. and Susskind, L.: {Fluid dynamics, Chern--Simons theory and 
the quantum Hall effect}, {\it Int. J. Mod. Phys.} B {\bf 5} (1991), 2735-50.
\ref
Bellissard, J.: {K-theory of C$^*$-algebras in solid state physics},
in T.C. Dorlas, N.H. Hugenholtz, and M. Winnink, (eds.) 
{\it Statistical mechanics and field theory: mathematical aspects}
Lecture Notes in Physics 257, Springer Verlag, Berlin-New York, 1986. 
\ref
Berest, Yu. and Wilson, G.: {Automorphisms and ideals of the Weyl algebra} 
{\it Math. Ann.} {\bf 318} (2000) 127-147.
\ref
Born, M. and Jordan, P.: {Zur Quantenmechanik}, {\it Z.f. Physik} {\bf 34} 
(1925), 858-888.
\ref
Cannings, R.C. and Holland, M.P.: {Right ideals in rings of differential 
operators}, {\it J. Alg.} {\bf 167} (1994), 116-141.
\ref Carey, A.L. and Langmann, E.: {Loop groups, anyons and the 
Calogero--Sutherland model}, {\it Commun. Math. Phys.} {\bf 201} (1999), 1-34.
\ref 
Connes, A. {\it Noncommutative geometry}, Academic Press, San Diego, 1994.
\ref
Connes, A.,Douglas, M.R.  and Schwartz, A.S.: {Noncommutative geometry and 
matrix theory}, {\it J. High Energy Phys.} {\bf 2}, Paper 3 (1998), 
hep-th/9711162. 
\ref
Corwin, L.: {Tempered distributions on the Heisenberg groups whose 
convolution with Schwartz class functions is Schwartz class}, {\it J. Funct. 
Anal.} {\bf 44} (1981), 328-347.
\ref
Dirac, P.A.M.: {The fundamental equations of quantum mechanics}, {\it Proc. Roy. 
Soc. A} {\bf 109} (1926), 642-653.
\ref
Douglas, M.  and Nekrasov, N.: {Noncommutative field theory}, 
{\it Rev. Modern Phys.} {\bf 73} (2001), 977-1029, hep-th/0106048.
\ref
Fr\"ohlich, J.  and Studer, U.M.: {Gauge invariance and current algebra 
in nonrelativistic many body theory}, {\it Rev. Mod. Phys.} {\bf 65} (1993), 
733-802. 
\ref 
Hannabuss, K.C.: {Noncommutative twistor space}, {\it Lett. Math. Phys.} 
{\bf 58} (2001), 153-166.
\ref
Harvey, J.A.: {Topology of the gauge group in noncommutative gauge 
theory}, hep-th/0105242.
\ref
Howe, R.: {$\theta$-series and invariant theory}, Proc. Symp. in Pure Math.
{\bf 33(i)}, {\it Amer. Math. Soc.} Providence Rhode Island, 1979.
\ref
Hellerman,S.  and van Raamsdonk, M.: {Quantum Hall physics equals 
noncommutative field theory}, {\it J. High Energy Phys.} {\bf 10} (2001), 
Paper 39, hep-th/0103179.
\ref
Kapustin, A., Kuznetsov, A. and Orlov, D.: {Noncommutative instantons and 
twistor transform}, {\it Commun. Math. Phys.} {\bf 221} (2001), 385-432, 
hep-th/0002193.
\ref
Kazhdan, D., Kostant, B.  and Sternberg, S.: {Hamiltonian group actions and 
dynamical systems of Calogero--Moser type}, {\it Commun. Pure and Appl. Math.} 
{\bf 31} (1978) 481-507.
\ref
Kouakou, K.M.: {\it Isomorphismes entre alg\`ebres d'op\'erateurs 
diff\'erentielles sur les courbes alg\`ebriques affines}, Th\`ese, Lyon 1994.
\ref
Lieb, E., Seiringer, R. and Yngvason, J.: {Bosons in a trap: 
A rigorous derivation fo the Gross-Pitaevskii energy functional}, 
{\it Phys. Rev. A} {\bf 61} (2000) 043602-1--043602-13, math-ph/9910033.
\ref
Littlewood, D.E. and Richardson, A.R.: {Group characters and algebra}, 
{\it Philos. Trans. R. Soc. A} {\bf 233} (1934), 99-141.
\ref 
Macdonald, I.G.: {\it Symmetric Functions and Hall polynomials}, 2nd edition
Clarendon Press, Oxford, 1995.
\ref 
Nekrasov, N.  and Schwartz, A.: {Instantons on noncommutative $\real^4$ 
and $(2,0)$ superconformal six-dimensional theory}, {\it Commun. Math. Phys.} 
{\bf 198} (1998), 689-703, hep-th/9802068.
\ref
Olshanetsky, M.A.  and Perelomov, A.M.: {Completely integrable Hamiltonian 
systems connected with semi-simple Lie algebras}, {\it Invent. Math.} {\bf 31} 
(1976), 93-108.
\ref
Polychronakos, A.: {Integrable systems from gauged matrix models}, 
{\it Phys. Lett. B} {\bf 266} (1991), 29-34.
\ref
Polychronakos, A.: {Quantum Hall states and matrix Chern--Simons theory}, 
{\it J. High Energy Phys.} {\bf 4} (2001) Paper 11, hep-th/0103013.
\ref
Seiberg, N.  and Witten, E.: {String theory and noncommutative geometry},
{\it J. High Energy Phys.} {\bf 3}, Paper 32 (1999), hep-th/9908142. 
\ref
Susskind, L.: {The quantum Hall fluid and noncommutative
Chern--Simons theory}, hep-th/0101029.
\ref
Wilson, G.: {Collisions of Calogero--Moser particles and an adelic 
Grassmannian} {\it Invent. Math.} {\bf 133} (1998), 1-41.

\bye